# Short Wave Infrared Reflectance Investigation of Sites of Palaeobiological interest: Applications for Mars Exploration

ADRIAN BROWN[1], MALCOLM WALTER[1], AND THOMAS CUDAHY[1,2]

[1] Australian Centre for Astrobiology, Macquarie University, NSW 2109, Australia, corresponding email: abrown@els.mq.edu.au

[2] CSIRO Division of Exploration and Mining, ARRC Centre, 26 Dick Perry Ave, Technology Park, WA , 6102 Australia

ABSTRACT

Rover missions to the rocky bodies of the solar system and especially to Mars require lightweight, portable instruments that use minimal power, require no sample preparation and provide suitably diagnostic mineralogical information to an Earth-based exploration team. Short Wave Infrared (SWIR) spectroscopic instruments such as the Portable Infrared Mineral Analyser (PIMA) fulfill all these requirements. We describe an investigation of a possible Mars analogue site using a PIMA instrument.

A survey was carried out on the Strelley Pool Chert, an outcrop of stromatolitic, silicified Archean carbonate and clastic succession in the Pilbara Craton, interpreted as being modified by hydrothermal processes. The results of this study demonstrate the capability of SWIR techniques to add significantly to the geological interpretation of such hydrothermally altered outcrops. Minerals identified include dolomite, white micas such as illite-muscovite and chlorite. In addition, the detection of pyrophyllite in a bleached and altered unit directly beneath the succession suggests acidic, sulfur-rich hydrothermal activity may have interacted with the silicified sediments of the Strelley Pool Chert.

# INTRODUCTION

Short Wave Infrared (SWIR) reflectance spectroscopy has gained recognition in the exploration and mining community due to its speed, simplicity and ability to characterize alteration zones around ore bodies. Recent SWIR studies (Huston *et al.*, 1999; Thompson *et al.*, 1999; Yang *et al.*, 2000; Herrmann *et al.*, 2001; Yang *et al.*, 2001; Bierwirth *et al.*, 2002; Thomas and Walter, 2002) have emphasized the instruments ability to detect alteration minerals such as white micas and chlorites. Each of these studies was conducted using the Australian-built Portable Infrared Mineral Analyser (PIMA) instrument.

To test the SWIR technique in a geologic setting comparable to the ancient flood basalts of Mars, fieldwork was undertaken in the arid, 3.5 Ga Pilbara Craton of Western Australia. Over 250 SWIR spectra were obtained by a hand-held "PIMA II" spectrometer at an outcrop of a heavily silicified, stromatolitic Archean carbonate-chert succession to simulate the investigation of such an outcrop by a rover on Mars.

To simulate the rover responding to commands from an Earth-based exploration team, spectra were taken only on the accessible weathered surface of the outcrop, and care was taken to target mineral assemblages that could be distinguished visually. This simulated a situation whereby a remote command team would possess only panoramic color images from the rover to select locations for collection of spectra.

## GEOLOGICAL SETTING

The North Pole Dome (NPD) in the East Pilbara Granite Greenstone Terrane (Van Kranendonk, 2000), is a structural dome of bedded, dominantly mafic greenstone sequences (the Warrawoona Group) that dip gently away from a central monzogranite. The monzogranite has been interpreted as a syn-volcanic laccolith, a product of diapiric uprise and consanguineous magmatism (Van Kranendonk, 2000). Minor occurrences of felsic volcanic rocks are interbedded with the greenstones, and these are capped by cherts that indicate hiatuses in volcanism (Van Kranendonk, 2000).

A stratigraphic column of the Warrawoona Group is presented at Table 1. Stromatolite and putative microfossil occurrences have been documented within the Warrawoona Group at three distinct stratigraphic levels – within the Dresser Formation, Apex Chert and Strelley Pool Chert (Dunlop *et al.*, 1978; Walter *et al.*, 1980; Awramik *et al.*, 1983; Lowe, 1983; Schopf, 1993; Ueno, 1998; Hofmann *et al.*, 1999; Van Kranendonk, 2000; Ueno *et al.*, 2001a; Ueno *et al.*, 2001b; Van Kranendonk *et al.*, 2003). The North Pole Dome has been interpreted as an early setting for life (Groves *et al.*, 1981; Buick, 1990).

<< insert Table 1 here >>

The rocks of the Warrawoona Group are dominated by thoelitic and komatiitic volcanic successions, which have been suggested as an analogue for the flood basalts of Mars (Baird *et al.* 1981; Reyes and Christensen, 1994; Mustard and Sunshine, 1995; Christensen *et al.*, 2000). Although the Earth and Mars have experienced vastly different weathering environments in the past 3.5 billion years, the ancient age of the Archean rocks of the North Pole Dome makes them a compelling analogue for similarly aged parts of the southern highlands on Mars. The preservation state of the Warrawoona group is generally excellent, in contrast with other Early Archaean terranes. Metamorphism has not exceeded greenschist facies throughout the North Pole Dome, and is most commonly at prehnite-pumpellyite facies (Van Kranendonk, 2000). The presence of putative microfossils and stromatolites at the North Pole Dome makes it an ideal test bed for Astrobiological techniques designed to examine primitive fossilized life.

A number of contributions have been made regarding the depositional setting of the Warrawoona Group. Some researchers have suggested a shallow marine environment based on the presence of pillow basalts, sedimentary analysis and REE systematics (Buick and Dunlop, 1990; Van Kranendonk *et al.*, 2003), whereas others have suggested an Archean mid-ocean ridge (MOR) due to geochemical analysis of basaltic successions and perceived similarities between hydrothermal alteration patterns at the NPD and in modern MORs (Kitajima *et al.*, 2001). The most recent interpretations invoke a caldera-like environment on an oceanic plateau based upon interpretation of volcanic conduits in the north of the NPD (Van Kranendonk and Pirajno, *in press*).

Barley (1984) suggested that silicification and carbonate-chlorite alteration of North Pole Dome greenstones were the result of low temperature, low pressure ("epithermal") hydrothermal alteration. Silicification associated with epithermal systems in early Archean greenstone terrains is almost ubiquitous (De Wit *et al.*, 1982; Gibson *et al.*, 1983). It has been proposed that silicification in the Dresser Formation and Strelley Pool Chert occurred very early after the deposition of a shallow subaqueous to subaerial evaporite sequence (Lowe, 1983; Buick and Dunlop, 1990). The process of early silicification is critically important for preservation of delicate biogenic structures such as microbial mats.

The Trendall Locality, within the Strelley Pool Chert in the south-west part of the North Pole Dome, contains a silicified carbonate-clastic succession with well-preserved stromatolites (Hofmann *et al,* 1999.). The origin of these stromatolites is controversial due to their ancient age and apparent lack of preserved microfossils. They are considered, by some researchers, to indicate the presence of a primitive biosphere at 3.4 Ga (Van Kranendonk *et al.,* 2003). Although early researchers provided evidence for normal marine stromatolite precipitation (Hofmann *et al.*, 1999; Van Kranendonk *et al.,* 2003), Lindsay et al. (2003) suggest the stromatolites were abiotic and formed by direct carbonate precipitation from syn-depositional hydrothermal activity. It is considered by the authors that the three-dimensional conical shape of the laminae (see Figure 2g) favors a biotic origin for these enigmatic features.

In any case, the Trendall locality has been affected by hydrothermal activity, either syn-depositional, post-depositional or both. We adopt the view that hydrothermal alteration has mediated the deposition of at least some parts of the SPC as a working hypothesis in this study.

## METHODOLOGY

Reflectance spectroscopy utilizing the SWIR region of the electromagnetic spectrum, (1.3-2.5 microns) often exploits absorption bands due to hydroxyl ($OH^-$) ions bonding with nearby cations in the crystal lattice. Minerals that contain hydroxyl ions are commonly associated with aqueous alteration. Such minerals are commonly formed by addition of hydroxyl ions to their crystal structure in hydrothermal systems, where hot (>50$^O$ C) water entraining solutes passes through rock pores or fissures.

Figure 1 displays the SWIR spectra of several alteration minerals typical of hydrothermal systems. The absorption bands around 2.2 micrometers are caused by a combination of the $v2$ fundamental stretching vibration mode of the $OH^-$ hydroxyl ion with the Al-OH bending mode in the crystal lattice of each mineral (Hunt, 1979). The central wavelength of the absorption band varies slightly due to the type of cation (for example $Fe^{2+}$ or Mg substituting for Al) ionically bonded to the hydroxyl ion. This spectral characteristic allows the determination of relative proportions of Mg or $Fe^{2+}$ to Al in white micas like muscovite-phengite. Another significant absorption band in the SWIR region is the carbonate ($CO_3^{2-}$) vibration mode at 2.32 micrometers, which also slightly changes wavelength with varying amounts of Ca and $Fe^{2+}$ to Mg in magnesite, dolomite and calcite (Gaffey, 1986).

<<insert Figure 1 here

caption:

SWIR spectra of common alteration minerals. Courtesy of ISPL (www.intspec.com).

>>

Table 1 gives a partial list of alteration minerals discernable using SWIR spectra, as well as details regarding their common mode of occurrence (Thompson *et al.,* 1999).

<< insert Table 1 here >>

Mapping alteration systems surrounding ore bodies has long been a pursuit of economic geologists (Meyer and Hemley, 1967). By recognizing distinctive mineralogies that typically form zones within hydrothermal alteration systems, suitable "vectors to ore" can be determined (Galley, 1993). For example, the presence of white mica or sericite is often the result of the breakdown of plagioclase when an igneous rock is hydrothermally altered. By mapping the occurrence of white mica, hydrothermal veins or alteration zones fluid flow can be delineated. Airborne SWIR studies are a particularly effective tool for regional mapping of white mica or sericite veins (Cudahy *et al.*, 2000; Brown, 2003a), and correlation on the ground is possible using hand held, or rover-mounted, instruments such as the PIMA.

The PIMA instrument measures reflected light from an internal light source at wavelengths between 1.3 and 2.5 micrometers. The instrument must be in direct contact with the sample during analysis, since it uses an internal light source for illumination. Samples may be unprepared outcrop (as in this study), rock chips or crushed powders in a Petri dish. The instrument integrates the reflected light from a small region of approximately 10mm diameter in front of the detector.

The bandwidth of the PIMA is approximately 7 nm and the spectral sampling interval is 2 nm, though its spectral resolution in the SWIR is closer to 8 nm. The unit has a Signal to Noise Ratio (SNR) of between 3500 and 4500 to 1 (www.intspec.com). Measurements typically take one to two minutes to acquire, depending on the selected integration time. Following each spectrum collection cycle, the instrument automatically carries out a reflection calibration against a known internal standard which is contained within the PIMA. Wavelength calibration must be carried out manually on a periodic basis by visual comparison of the spectrum of the internal calibration material with a previously established standard spectrum of the same material. The instrument operates from a 12V NiCd battery. Internal temperature and battery status are measured and reported to an attached WinCE palm computer or laptop PC. Spectra from the PIMA instrument can be downloaded for further analysis under programs such as Microsoft Excel or "The Spectral Analyst" (TSA).

PIMA instruments have been manufactured by Integrated Spectronics, Pty Ltd. since 1991. Approximately 250 units are in service throughout the world. Two models were produced; the PIMA II and the upgraded PIMA SP model. Further PIMA models are under development (T. Cocks, pers. comm.).

In this study, "The Spectral Analyst" (TSA, version 4) program was used to obtain an automatic assessment of the acquired SWIR spectra. The TSA automatic assessment program is designed to compare the acquired spectrum to a library of pure endmember spectra, and a proprietary algorithm identifies a primary and secondary mineral present in the analysed sample. If the spectrum lacks sufficient infrared-active features for the automatic assessment program to identify the phase(s), the rock is declared "aspectral". This automated procedure simulates a scenario where a rover could independently determine the minerals present in rocks at a landing site and report the results of its survey to a remote science team. The spectra were then manually inspected using "The Spectral Geologist" software and visual assessments were used to interpret the mineral assemblage according to shape and wavelength minima of detected absorption bands. This technique allowed the automatic spectral identification techniques of TSA to be tested and assessed for accuracy.

TSA uses a database of 500 samples of 42 SWIR active library (or "endmember") minerals (Pontual, 1997), which are used to determine the closest match for unknown sample spectra. Similar databases have been developed for reflectance spectroscopy techniques by the USGS (Clark *et al,* 1990).

Reflectance spectra are often characterized by their albedo, which is the overall reflectance response of a sample. 'Dark' samples have a low albedo. SWIR reflectance spectra display a spectral 'continuum' that is shaped like an inverted hull. The continuum or hull shape may be relatively flat or significantly curved, and this may help in identifying certain samples, but most often it interferes with identification of absorption bands which are diagnostic of certain minerals. Many software products, including TSA, provide a method for removal of the spectral continuum to better identify spectral absorption bands. This normalization method, termed "convex hull removal" (Clark *et al.*, 1987), has been employed in this study when visually investigating spectra. Spectra displayed within this report have all been processed with the convex hull removal procedure.

## RESULTS

This survey covered a 50m x 30m section of outcrop at the Trendall Locality, an outcrop of the Strelley Pool Chert. To support the collection of spectra, 1500 digital photographs were taken of the outcrop; these are available online at http://aca.mq.edu.au/abrown.htm. Figure 2 displays a geological map interpreted from these photos of the Trendall Locality.

<< insert Figure 2 here

caption:

Geological map of the Trendall Locality re-created from digital images of the outcrop. Original geological map of the outcrop by Van Kranendonk and Hickman (2000).

>>

A visual assessment was made of rock types at the Trendall locality before spectra were taken. This simulated the classification of rocks in the vicinity of a robotic rover by remote scientists using panoramic images. Similar panoramic images were used intensively in NASA's Pathfinder (McSween *et al.,* 1999) and MER (Squyres, 2003) missions. On the basis of this assessment, rocks were broadly categorized according to color and texture, as indicated in Table 2. Examples of units present at the locality are shown in Figure 3.

<< insert Table 2 here>>

<< insert Figure 3 here

caption:

Images of example outcrop of units described in this paper. Geological hammer for scale. a. i.) wide laminae black and white chert, a. ii.) fine laminae black and white chert, b.) basalt, c.) planar carbonate with stromatolites, d.) mudstone, e.) boulder conglomerate, f. i.) bulbous pyrophyllite schist in white groundmass, f. ii.) foliated pyrophyllite shist, g.) siliceous planar laminate displaying conical stromatollites.

>>

Following the broad categorization based on visually distinct rock type, the area was surveyed with the PIMA spectrometer in the manner a rover might examine the outcrop. This was done by taking a number of spectra of each identified unit. At least 7 spectra were taken of each rock unit, sufficient to determine its broad spectral characteristics, by visual recognition of absorption bands present in each measurement of a unit and identifying any significant changes between spectra within the unit.

<< insert Table 3 here>>

The spectral features of each unit are discussed below. Apart from the overlying basalt unit and the underlying pyrophyllite schist unit, all units are part of the Strelley Pool Chert.

BASALT. The overlying Euro Basalt is characterized by a light brown weathering rind, and a green chlorite-rich mineral assemblage. It is fine grained and largely eroded at this location. Many samples are present as partially buried boulders rather than as competent outcrop.

The TSA analysis of the PIMA spectra of the basalt unit consistently identified Mg-chlorite within the unit (Table 3). An example spectrum is shown in Figure 4. The chlorite was identified by its diagnostic absorption band centered at 2.25 micrometers, lack of a feature at 1.55 micrometers (typical of epidote), and associated chlorite features at 2.0 and 2.33 micrometers (McLeod *et al.*, 1987).

<< insert Figure 4 here

caption:

Convex hull removed example spectra of units found at the Trendall Locality.

>>

MUDSTONE. The mudstone unit is characterized by a grey fine-grained cherty groundmass, containing millimeter sized white clasts, commonly showing a sugary texture indicative of silicification. It lies near the top of the Trendall locality, among strata that have been classified as part of a clastic sequence based on previous research at this locality (Van Kranendonk and Hickman, 2000).

The spectra from the mudstone unit showed variable signatures though most displayed a symmetric absorption band at 2.2 microns, diagnostic of illite-muscovite. They also demonstrated variation in the amount of bound water, characterized by the absorption band at 1.91 microns. Illite generally has a deeper absorption band at 1.91 microns than muscovite. A typical spectrum of the mudstone unit is shown in Figure 4. Visual interpretation of the mudstone spectra suggests a white mica is definitely present within the unit, a conclusion based on the regular appearance of the Al-OH absorption band at 2.2 microns.

PEBBLE CONGLOMERATE. The pebble conglomerate unit is characterized by a grey-green coarse-grained groundmass, containing large (up to 2cm) ovoid clasts. The clasts are often white and sometimes derived from underlying units, primarily black chert and mudstone. This unit is also part of the clastic succession at the Strelley Pool Chert (Van Kranendonk and Hickman, 2000). The presence of rip up clasts of underlying units argues for a high energy subaqueous depositional environment.

Since the conglomerate is constituted by clasts from the mudstone unit (amongst others), it is not surprising that the spectra of the pebble conglomerate unit were similar to the mudstone unit.  It appeared, however, that the pebble conglomerate contained less water, an interpretation based on the presence of a weaker 1.9 micrometer absorption band, and the fact that this band was characterized by a flatter hull shape. The unit was variably chloritised - strong chlorite absorption bands at 2.25 and 2.33 microns were present in some samples, but absent in the majority of samples (Table 3).

BOULDER CONGLOMERATE. This unit consists of large (up to 5-10cm) clasts, commonly of black chert, but also including black and white layered chert clasts and planar layered carbonate. The clasts often display red to purple staining due to the presence of iron oxides.

This unit displays overall low reflectance (typically < 20%). Most spectra showed water absorption at 1.9 micrometers that included bound water (centered at 1.915 micrometers) and unbound water (centered at 1.93 micrometers). Unbound water is typical of free, adsorbed and trapped water (as expected for chert with ~1% $H_2O$) whereas bound water can be associated with minerals like illite and smectites (Aines and Rossman, 1994). Additional mineral absorptions of the spectra of this unit included symmetric bands at 2.2 micrometers due to the presence of illite-muscovite and coupled 2.25 and 2.33 micrometer absorptions indicative of chlorite.

TSA could not identify any characteristics in a proportion of the spectra from this unit which contained black chert clasts, and they were designated aspectral (Table 3). That is not surprising given the fact that chert is inactive in the short wave infrared. TSA suggested the presence of opaline silica in one sample, as shown by the presence of a broad Si-OH absorption band at 2.215-2.25 micrometers.

The presence of lithified planar laminate carbonate, black chert and quartz clasts within this conglomerate (see Figure 3e) confirms that it is older than the planar carbonate unit and some of the black chert had been lithified prior to deposition of this unit. The lack of planar siliceous clasts suggests that the silicification of the planar carbonate occurred after the conglomerate was formed.

BLACK CHERT. The black chert unit consists of massive microgranular black chert. The black color is due to minor amounts of kerogen within the chert (Kato and Nakamura, 2003).

At the Trendall Locality, massive black chert occurs typically as smooth, fine grained chert layers 5-10 cm thick when bedding conformable (often interleaved with other units, such as the mudstone unit), or as massive crosscutting veins 30-50 cm thick when oriented approximately normal to the stratigraphic layering. For the purposes of this study, the black chert laminations associated with white quartz layers were analyzed separated and grouped as the "black and white chert" unit (discussed below).

The PIMA spectra of the black chert unit were extremely dark (generally <10% reflectance) and showed much weaker (if present) water and mineral absorption bands compared to the other units studied.  TSA assigned a majority of the spectra as aspectral (Table 3). The spectra displayed low SNR due to their extremely low albedo. In low albedo situations, the imposition of deviations due to instrument related noise upon a small signal makes identification of absorption bands extremely difficult, even after convex hull removal.

BLACK AND WHITE CHERT. The black and white chert unit is present in two forms (see Figure 3a). The fine layered form is restricted to a relatively small area of the outcrop in the center of Figure 2 and consists of microgranular black chert interstratified at a millimeter scale with microgranular white chert, in parallel laminae displaying flat bottoms and curved tops. The morphology of this unit suggests direct periodic sedimentary deposition in a quiet environment.  The coarse layered form is laminar, and the width of laminae can range up to 5-10cm. The wide laminae are sub-parallel to other units at the locality, and often appear to radiate from the vertical black dykes described above. For this reason, the wide laminae black and white chert is likely a result of hydrothermal injection. Both forms of black and white layered chert display similar smooth microgranular texture. In most cases, the black chert surrounds the white chert. Both forms are generally found below the clastic upper units and above the carbonate and siliceous planar laminar units at the Trendall Locality.

The PIMA spectra of both forms of the black and white cherts could not be separated and are reported together here. The black and white chert showed higher albedo than the spectra of the black chert unit, though the water and mineral absorption bands were generally of similar relative intensity. TSA classified most of the spectra as aspectral (Table 3) even though absorptions at 2.2 and 2.3 micrometers were apparent. These small features are most likely due to small amounts of white mica and carbonate.

QUARTZ. Where quartz occurs in small (less than 4cm wide) veins, and displays irregular surficial textures (unlike the smooth chert units) it was identified as a separate unit for the purposes of this study. The veins were not mapped, and are not shown in Figure 2, due to their small size. However, they all occurred within the black and white chert unit and in most cases represent vug filling quartz that grew into cavities.

The PIMA spectra of these quartz veins showed a much stronger development of the unbound water absorption band centered at 1.93 microns, as shown by a representative spectrum in Figure 4. The broad and more defined character of this band contrasts with the weak and generally narrow bands due to bound water typical of the black cherts discussed above. Some quartz spectra were very dark with albedos ~10%, similar to those of the black cherts. A small number of the samples were identified by TSA as opaline silica (Table 3), based on the presence of an absorption band at 2.25 micrometers. Some spectra showed symmetric Al-OH absorption bands, possibly related to illite-muscovite. However, most of the samples were classified by the TSA program as aspectral. The presence of opal is most likely due to late stage weathering, evidenced by the sinuous, crosscutting and vug filling nature of the quartz veins.

RADIATING CRYSTAL SPLAYS. Radiating crystal splays were identified by earlier researchers (Van Kranendonk and Hickman 2000) and were likened by them to beds of aragonite deposited in modern day travertine (eg. Jones *et al,* 1997.). Recent trace element geochemical studies suggest the unit represents a dolomitized replacement of radiating crystal fans, interpreted as secondary crystal growth below the sediment-water interface (Van Kranendonk *et al.,* 2003).

Although visibly silicified, the spectra of the unit identified dolomite within the unit due to the presence of a wide absorption band at 2.31 micrometers.

PLANAR LAYERED CARBONATE. Planar laminated carbonate beds with pervasive conical stromatolites (Hofmann *et al.,* 1999) are present in the south-east part of the outcrop (bottom left of Figure 2). At point A on Figure 2, the unit is laterally crosscut by a sub-vertical black chert, which divides well preserved planar laminated carbonate from the siliceous planar laminated unit.

The spectra of the planar layered dolomite displayed a bimodal nature (Figure 5). Where stromatolite laminae had been preserved, the presence of Al-OH bonding indicative of kaolin was revealed by a deep asymmetric absorption at 2.2 microns. This was accompanied by a relatively weak carbonate absorption band at 2.31 microns. Where there was no preserved stromatolite laminae, the spectra lacked an absorption band at 2.2 microns and displayed a relatively strong carbonate absorption at 2.31 microns.

<< insert Figure 5 here

Caption:

a.) Example spectra taken from preserved stromatolite laminae (above) and carbonate regions (below). b.) Graph displaying antipathetic relationship of abundance of carbonate and kaolin within planar carbonate region. See text for explanation.

>>

As shown in the graph in Figure 5b, the antipathetic nature of the strength of the absorption bands at 2.2 and 2.31 microns in the planar carbonate layer demonstrates that a kaolin mineral is obscuring the carbonate beneath. The kaolin is most likely a weathering product or desert varnish caught in the closely spaced laminae following Aeolian movement. This hypothesis was strengthened when the fresh underside of a hand sample was examined with the PIMA and found to be dolomite with no trace of kaolin, even on the preserved stromatolitic laminae, thus showing the kaolin was not an original part of the stromatolite laminae. This process of desert varnish trapping by stromatolitic laminae may be relevant on the wind swept plains of Mars.

The nature of the weathering or varnish, and its exclusive distribution upon the exposed preserved laminae, which sit proud of the pure carbonate, is not entirely clear, but may provide one method for detecting preserved stromatolitic laminae in outcrops of Martian rocks.

Beneath the planar layered carbonate lies a region devoid of stromatolite laminae. Irregular sinuous veins of quartz penetrate this region in a sub-vertical orientation, and terminate at a thin layer of black chert just below the lowest stromatolite laminae. The presence of unlaminated carbonate beneath the stromatolite layer suggests carbonate deposition commenced prior to pervasive stromatolite growth. The origin of the quartz veins is not entirely clear, however they could be related to later hydrothermal heating.

The radiating crystals region discussed earlier may be analogous to the unlaminated carbonate region due to their similar orientation (beneath the planar layered carbonate and siliceous units) and lack of stromatolitic features.

SILICEOUS PLANAR LAMINATE. This unit was visually assessed as the silicified portion of the planar carbonate layer based on the presence of stomatolitic laminae of similar vertical and horizontal dimensions. The dominant color of the unit is dark grey to white.  A smooth microcrystalline siliceous texture exists between the stromatolitic layers.

Horizontally, the siliceous planar laminate is separated from the planar carbonate regions by a vertical black chert dyke (point A on Figure 2). The siliceous planar laminate also occurs above the carbonate planar unit, and again is separated by a horizontal black chert dyke. The presence of the black dykes surrounding the carbonate layer suggests that the silicification process was impeded by the black chert, therefore the black cherts were in position before the silicification event occurred. This suggests that although the vertical black dykes at the Trendall Locality may have been responsible for transportation of some hydrothermal fluids, such as those containing kerogen that contributed to the horizontal wide layered hydrothermal injection black and white cherts discussed earlier, it is unlikely they were fluid conduits for the major silicification of the planar layered carbonates, which was probably a later event.

The partially silicified nature of this carbonate unit was confirmed by the PIMA spectra, many of which showed minor absorption bands around 2.31 micrometers indicative of dolomite. Several spectra identified the presence of white mica, due to the presence of an Al-OH related symmetric absorption band at 2.2 microns. About half the spectra were designated by the TSA program as aspectral. (Table 3).

Areas of intense red, brown and yellow discoloration occur sporadically within the generally white siliceous unit. Several spectra from this region of the outcrop showed the distinctive bands of goethite at approximately 1.65 and 1.75 micrometers. The sporadic nature of the discolouration suggests relatively recent weathering may be responsible for these features.

The PIMA spectra of the iron-oxide stained stromatolitic laminae showed weak evidence for kaolin with absorption bands centered at at 2.165 and 2.209 micrometers. Generally the spectra of the stromatolitic laminae were found to lack diagnostic absorption bands, but all displayed moderately high albedo.

PYROPHYLLITE SCHIST UNIT. Stratigraphically beneath and to the south east of the area depicted in Figure 2 lies a bleached brown to golden colored, heavily foliated schist that extends vertically for approximately 50m beneath the Strelley Pool Chert. There is no outcrop at the contact between the chert and schist units, however as the pyrophyllite schist unit converges on the base of the Strelley Pool Chert, the alteration style changes from platy, foliated brown to golden outcrop to smaller bulbous outcrop surrounded by a white fine-grained groundmass (see Figure 3f). The sample locations are shown in the inset of Figure 2.

The spectra taken of the bulbous and foliated forms of the pyrophyllite schist unit clearly showed the presence of pyrophyllite, indicated by a diagnostic symmetric and sharp absorption band centered at 2.165 micrometers, accompanied by a smaller shoulder absorption band at 2.195 microns. A representative example of this spectrum is shown in Figure 4. Pyrophyllite is a typical alteration mineral resulting from acidic hydrothermal fluids rich in sulfur (high-sulphidation) (White and Hedenquist, 1990). No clear linkage has been established between the high-sulphidation alteration beneath the SPC and the paragenesis of units within the SPC, however since no trace of pyrophyllite is found in the Euro basalt unit overlying the SPC, it is likely the SPC formed prior to the pyrophyllite schist unit and acted as a barrier to the later high sulphidation hydrothermal fluids. An alternative explanation is that the SPC units were laid down on a partially eroded phyrophyllite schist unit, however no evidence for an unconformity has been reported between the units.

## VARIATION IN OCTAHEDRAL AL OF WHITE MICA

Tschermak substitution in phyllosilicate minerals, where two Al atoms substitute for a Si atom and either a Fe or Mg atom in the crystal lattice [$Al_2Si_{-1}(Fe,Mg)_{-1}$], can be recognized in the SWIR by analysis of the Al-OH absorption band position at around 2.2 micrometers (Duke, 1994). An increase in the Al content in a crystal structure containing Al-OH is reflected by a decrease in the wavelength of the 2.2 micrometer absorption band. The wavelength variations typically range from 2.217 to 2.197 micrometers.

All units displaying the Al-OH absorption band around 2.2 microns were analysed for variations in the wavelength of the the band minima. No significant Tshermak variations were found within the units at the Trendall Locality, however small variations were found between units. Table 4 summarises the differences in average position of the Al-OH absorption bands within Al-OH bearing units.

<< insert Table 4 here >>

The lack of Tschermak variation within units at the Trendall locality does not rule out variations in the same units some distance from this outcrop. These results suggest that at the scale of this outcrop, approximately 50m across, there were no significant temperature variations and resultant Al abundance within hydrothermal white mica.

## GEOCHEMICAL ANALYSIS

To support the PIMA mineral identification in the samples studied, XRD spectra were taken of selected samples from the Trendall locality. Due to the historical importance and current state of preservation of the site, sampling is discouraged by the Geological Survey of Western Australia and so only limited sampling was undertaken. The analysis was carried out by Sietronics Pty Ltd using a Bruker AXS D4 X ray diffractometer, and CSIRO using a Phillips PW 1050 diffractometer using CuK α radiation.

Two samples of the planar carbonate unit were analyzed. The results showed the presence of quartz and dolomite. No other minerals were detected. Rietveld analysis (Rietveld, 1969) was carried out using the "Siroquant" program (Taylor, 1991). The resultant data was used to estimate the sample composition to be 45 wt. % quartz, 55 wt. % dolomite. The XRD analysis thereby supported the SWIR detection of dolomite in the planar carbonate unit. One sample of the pyrophyllite schist unit was analyzed, and the XRD data showed positive for pyrophyllite and quartz. Samples of the black and white chert, mudstone and silicified chert were also analyzed and the data indicated the presence of quartz and no other minerals.

## DISCUSSION

Previous studies have demonstrated the ability of SWIR techniques to map alteration minerals in hydrothermally altered terrain (eg. Thompson *et al.*, 1999). This study has shown the utility of using a SWIR instrument to identify mineral phases indicative of hydrothermal alteration in a hydrothermally altered, silicified, stromatolitic carbonate and clastic succession.

Highly silicified environments are typical of Archean greenstone terrains that have undergone moderate hydrothermal alteration (Gibson *et al.*, 1983; Duchac and Hanor, 1987;

Van Kranendonk and Pirajno, in press). Archean greenstone terrains often contain cherts that preserve textures and fabrics indicative of past biological activity such as stromatolites (De Wit *et al.,* 1982). Some workers have indicated that chert deposits may be ideal places to search for fossilized life on Mars (Walter and Des Marais, 1993).

Similarities between hydrothermal events in the Pilbara and Mars. It has been postulated that impact craters (Newsom *et al.,* 2001) and sites of gully formation (Brakenridge *et al.*, 1985; Gulick 1998) may be sites on Mars where hydrothermal waters may have penetrated the martian regolith. Sites of hydrothermal activity on Mars may display silicification and cherty sediments such as those found at the Trendall Locality, and thus be ideal for the preservation of microbial mats as laminae (Walter and Des Marais, 1993).

Although SWIR instruments are limited in their ability to examine low albedo black cherts, we have shown that the SWIR technique can identify minerals such as carbonates and phyllosilicates and can be used to characterize the degree and nature of alteration of such minerals, a critical precursor to the localization and discovery of life or preserved biogenic fossils.

In the course of a typical rover mission on Mars, and particularly if a rover were to encounter a site of Astrobiological interest on Mars, the minerals present at the site would need to be quickly surveyed to localize areas where further, more time-consuming analyses could be carried out. A SWIR instrument such as the PIMA could serve this role for a rover operating in such a reconnaissance mode.

In preparation for the 2003 Mars Exploration Rover missions, NASA science and engineering teams used an instrument called IPS, which is similar to PIMA, on its evaluation rovers (Haldemann *et al.,* 2002). This instrument differs from PIMA in that the instrument does not have to be in contact with the rock in order to obtain a spectrum. The light of the sun is used by IPS as an illumination source. The instrument was found to add considerably to the geologists' ability to characterize mineralogically distinct geological sequences in a rover-remote science team scenario (Jolliff *et al.*, 2002).

This study shows that automatic rock classification using a SWIR reflectance spectroscopy instrument such as PIMA (Pedersen *et al.*, 1999; Moody *et al.*, 2001; Gulick *et al.*, 2003) would prove useful in future extraterrestrial robotic missions. Mineral identification algorithms such as those used by TSA may prove useful in such a system, however this study has highlighted the difficulty of automatic recognition of low albedo reflectance spectra (such as those of black cherts), due to low achieveable signal to noise.

The ability of the PIMA to identify mineral modes makes them ideal to support elemental abundance instruments such as the Alpha Proton X-ray Spectrometer (APXS) (Golombek, 1998) or Laser Induced Breakdown Spectrometer (LIBS) (Wiens *et al.,* 2002). The minerals modes provided by a PIMA spectrum could be quantified using a standard normalization technique when combined with an elemental abundance measurement. Unusual hydroxyl bearing mineral assemblages, such as jarosite and saponite, can be detected unambiguously by the PIMA.

## CONCLUSIONS

This study has demonstrated the ability of SWIR reflectance spectroscopy to identify minerals in an Archean hydrothermally-altered environment. Minerals identified included carbonates, such as dolomite, white micas, such as illite-muscovite, opaline silica, clays, and chlorite. These identifications supplemented visual assessments and aided in the geological interpretation of the outcrop under study.

The SWIR spectroscopic method is ideal for mounting on a small rover for remote field analysis of collected samples. It is lightweight and has low power requirements, being able to be operated with a 12V battery. With an instrument such as the PIMA, calibration and spectrum collection can be completed in around two minutes. The availability of a large body of research and spectral libraries coupled with advanced algorithms in packages such as "The Spectral Analyst" make the technique a low-risk supplementary and reconnaissance instrument suitable for mineral detection of samples that require virtually no sample preparation. We have demonstrated its utility even in highly silicified environments.

The results of this study show that SWIR spectroscopy contributed to the classification and interpretation of the geology of the Strelley Pool Chert. Previous studies of this area have now been augmented with the following information from our analyses:

a.)     the presence and extent of carbonate and partially replaced dolomitic carbonate was confirmed,

b.)     preserved stromatolitic laminae at the Trendall Locatlity are covered by a kaolin -rich desert varnish, most probably trapped by the stromatolitic laminae, in contrast to the purely dolomitic nature of the carbonate groundmass, and c.)     pyrophyllite was detected in an extensively altered and foliated unit (pyrophyllite schist) beneath the silicified outcrop, which suggests that acidic high sulphidation hydrothermal alteration has taken place directly beneath the Strelley Pool Chert and may have interacted with the SPC unit.

## ACKNOWLEDGEMENTS

| Age (Ga) | Unit | Fossil assemblages |
|---|---|---|
| | Euro Basalt | |
| **3.458** | **Strelley Pool Chert** | **Well preserved conical stromatolites (Hofmann et al., 1999)** |
| | Panorama Formation | |
| | Apex Basalt | Microfossils within Apex Chert (Schopf, 1993) |
| 3.470 | Duffer Formation | |
| | Dresser Formation | Domical stromatolites and microfossils (Walter et al. 1980; Awramik 1983) |
| | Mt Ada Basalt | |
| 3.515 | Coonterunah Group | |

TABLE 1 – STRATIGRAPHIC COLUMN OF WARRAWOONA GROUP UNITS PRESENT AT THE NORTH POLE DOME (VAN KRANENDONK, 2000). DATES ARE DERIVED FROM U-PB ISOTOPES FROM ZIRCONS IN THE UNITS AND ARE ACCURATE TO APPROXIMATELY 3 MILLION YEARS FROM THORPE (1992).